\documentclass{ws-rv961x669}
\begin{document}

\chapter{The Third Cosmological Paradigm}         
\author{Michael S. Turner}        
\address{The Kavli Foundation and the University of Chicago\\
mturner@uchicago.edu\footnote{The author is Senior Strategic Advisor at the The Kavli Foundation and Rauner Distinguished Service Professor emeritus at UChicago.}}

\begin{abstract}
I begin by briefly discussing the first two cosmological paradigms, the hot big-bang model and $\Lambda$CDM.  In discussing the third paradigm, I focus on the issues it must address, what its aspirations should be, and how it might be initiated.  I end with a brief history of my collaborations with Frank Wilczek.
\end{abstract}

\section{The first epoch (circa 1920 to 1980):  the emergence of the Hot Big Bang paradigm}       
The Universe is big in both space and time and for much of human history most of it has been beyond the reach of our boldest ideas and most powerful instruments.  I mark the birth of modern cosmology at roughly 100 years ago.  Einstein had introduced general relativity, the first theory of gravity and spacetime capable of describing the Universe, and the first cosmological solutions had been found (e.g., the deSitter, Friedmann and Lema{\^ i}tre solutions as well as Einstein's static model).  At about the same time, George Ellery Hale and George Ritchey invented the (modern) reflecting telescope, and Hale moved astronomy to the mountain tops of California, first Mt. Wilson and later Mt. Palomar. With bold ideas and new instruments, we were ready to explore the Universe beyond our own Milky Way galaxy and begin to discover and understand the larger picture.

Hale's second big reflector, the 100-inch Hooker telescope, enabled Hubble to discover that galaxies are the building blocks of the Universe today and that the Universe is expanding, the signature of its big bang beginning.  While it took a few years to connect the solutions of general relativity to the observational data, by the early 1930's the basic big bang model was in place (and Hubble was on the cover of {\it Time} magazine).

The final element of the hot big bang model came with Arno Penzias and Robert Wilson's discovery of the cosmic microwave background (CMB) in 1964.\footnote{Radio astronomy, a field pioneered by physicists Karl Jansky and Grote Reber in the 1930s, was the first of many new windows on the heavens--beyond visible light--to be opened.}  While the idea of a hot beginning was introduced by George Gamow and his collaborators in 1948, to explain the non-equilibrium origin of the chemical elements,\footnote{In an odd twist, Gamow's version of big-bang nucleosynthesis was largely wrong; in fact, it was a description of what is known today as the r-process.  Hoyle, worried that his steady state model lacked the ability to explain the chemical elements, produced the modern version of the r-process in his landmark paper with the Burbidges and Willy Fowler.} the 1964 discovery was accidental, an interesting and oft-told story.

In 1972, years before Standard Model referred to the remarkable theory that describes quarks and leptons, Steven Weinberg coined the term ``Standard Model" for the hot big bang model and described it in his classic textbook \cite{SWeinberg}.  In brief, the model traces the Universe from a hot soup of hadrons at around $10^{-5}\,$sec through the synthesis of the light elements (largely $^4$He with traces amounts of D, $^3$He and $^7$Li) at a few seconds to the formation of neutral atoms and the last-scattering of CMB photons at around 400,000 years after the big bang to the formation of stars and galaxies.

The first paradigm laid out the basic architecture of the Universe -- expansion from a hot big bang beginning to a Universe filled with galaxies moving away from one another today.  General relativity, nuclear physics (for big bang nucleosynthesis, or BBN), and atomic physics (for the CMB) physics provided a strong theoretical foundation.  The triad of the expansion, the light-element abundances, and the blackbody spectrum of the CMB provided an equally strong observational foundation.

Allan Sandage summed up cosmology in 1970 as the search for two numbers, $H_0$ and $q_0$, the expansion rate of the Universe and the deceleration rate of the Universe respectively \cite{ASandage}.  It would take until 2000 and the completion of the Hubble Space Telescope Key Project to pin down $H_0$ to 10\% with a reliable error estimate:  $H_0 = 72 \pm 2 \pm 6\,$km/s/Mpc \cite{HSTKeyProject}, and as I will discuss later, $H_0$ is still a lively topic today.  As for the deceleration parameter, the Universe is actually accelerating; further, $q_0$ is not even measurable with both precision and accuracy, another story \cite{NebenTurner}.  Bottom line, $H_0$ continues to live up to its reputation as the most important number in cosmology, and $q_0$, whose inferred value is about $-0.55$ give or take 10\%, has fallen by the wayside.

There were also other important open issues in the first epoch.  Chief among them, precisely how structure formed and what happened during the first microsecond when the soup of hot hadrons would have been strongly-interacting and overlapping.

One last thought on this first period:  the sociology.  The field was largely the province of astronomers and only a handful at that (certainly less than 30).  There were the occasional physicists, e.g., Einstein, Gamow, and Tolman, who dabbled.  The discovery of the CMB began to bring in physicists:  Penzias and Wilson (radio astronomy physicists), Wagoner and Fowler (who with Hoyle carried out the first modern calculation of BBN \cite{WFH}) and the Princeton gang who interpreted the Penzias and Wilson discovery (Dicke, Peebles, Roll, and Wilkinson \cite{Dickeetal}).  This foreshadowed the large number of physicist-cosmologists we see today, that began coming in big numbers during the second epoch.


\section{The second epoch (circa 1980 to 20??): the emergence of the $\Lambda$CDM paradigm}
Exactly when and how the paradigm shift occurred is hard to precisely pin down and likely involved two steps.  According to Thomas Kuhn, the push behind a paradigm shift are anomalies that the current paradigm cannot explain or account for.  There is also the pull of a better theory.  Both elements were at play here.

Dark matter most clearly conforms to a Kuhnian shift.  Around 1980, through the work of Vera Rubin and others \cite{RubinDM}, it became clear that bulk of matter that holds together galaxies is not in the form of stars, but rather in extended haloes of non-luminous (dark) matter.  This, together with Zwicky's earlier work about dark matter in clusters of galaxies and Ostriker and Peebles's work on the theoretical need for massive halos to stabilize the disk structure of spiral galaxies, brought astronomers to the realization that there is much more to the Universe than meets the eye.  But what is it?

New ideas in particle physics that trace to the discovery of asymptotic freedom by Gross \& Wilczek and Politzer provided a powerful pull:  During the first microsecond after the big bang the Universe was comprised of a soup of weakly-interacting, point-like quarks, leptons, gauge and Higgs bosons of the Standard Model and probably other fundamental particles.  Weinberg's  ``hadron wall'' fell and  early-Universe cosmology was open for business!  Further, the convergence of the three coupling constants led to grand unified theories and a flood of cosmological consequences including baryogenesis, inflation, magnetic monopoles, and particle dark matter.  By 1990, particle dark matter had caught on, even among astronomers.\footnote{Interestingly, Vera Rubin was amongst the astronomers who was never a fan of particle dark matter.}

Equally important, the ``heavenly lab" can extend the reach of ground-based accelerators with its great variety of densities, temperatures, energies and magnetic fields that are not available in terrestrial labs.  Zel'dovich called the early Universe ``the poor man's accelerator".

The successes of these powerful ideas from particle physics at addressing big puzzles in cosmology has established a lasting and deep connection between the very small -- elementary particle physics -- and the very large -- cosmology.  The connection between quarks and the cosmos not only changed how we think about cosmology and the very vocabulary we use, but it also brought in a host of new players, theorists and experimentalists from both particle physics and nuclear physics, dramatically increasing the numbers of scientists engaged in cosmology one way or another.

The quarks/cosmos connection goes far beyond the simple fact of an early, quark-soup phase of the Universe.  The events that occurred during (and before) the quark-soup phase have shaped the Universe we see today:  from the baryon asymmetry and dark matter to the origin of the large-scale regularity of the Universe and the seeds for all the structure seen today.  

Beyond the influx of new players and new ideas from physics, important technological advances were occurring in astronomy -- the widespread use of CCD cameras that today exceed a gigapixel in size, the opening of additional windows on the Universe from infrared and UV to x-rays and gamma-rays, more powerful CMB instruments and experiments (e.g., HEMT and bolometer detectors and the use of interferometry).  And the exponentially-increasing power of computing has made possible numerical simulations of the Universe and the analysis of the ever-larger datasets being created.  

By the beginning of 21st century, a new kind of astronomy had been born -- digital surveys of large portions of the sky, both imaging and spectra.  By 1980, the total number of redshifts measured was around a thousand with the highest redshift galaxy having a redshift of less than 1!  The pioneering Sloan Digital Sky Survey (a collaboration of astronomers and particle physicists) imaged more than 100 million galaixies and measured redshifts for one million of them.  Today, galaxy redshifts extend to 10 and the number redshifts measured is approaching one billion (or about 0.5\% of the galaxies in our visible Universe).

Building upon the strong foundation of the hot big bang model, a new set of questions were asked were asked during the second epoch -- and by 2000 were mostly answered:
\begin{itemize}
\item{}  How did structure form and what is the origin for the seeds of that structure?
\item{}  Why is the Universe -- at least the observable Universe -- so regular, i.e., isotropic and homogeneous, on the largest scales?
\item{}  Why does the Universe contain only matter and not equal amounts of matter and antimatter?
\item{}  What is dark matter comprised of?
\item{}  What explains the early flatness of the Universe and is it is still flat today?
\item{}  Where did all the magnetic monopoles (and other nasty relics of the early phases of the Universe) go?
\item{} What explains the smallness or absence of a cosmological constant (in the form of quantum vacuum energy)?
\end{itemize}

Beginning with baryogenesis -- the theory of how $B$, $C$ and $CP$ violating processes in the early Universe allowed the evolution of a small excess of baryons over antibaryons -- ideas motivated by grand unification, supersymmetry and eventually superstrings gave rise to the basic elements of $\Lambda$CDM:  inflation, particle dark matter and dark energy.  By 1985, the inflation and cold dark matter (slowly-moving dark-matter particles) paradigm was guiding cosmology.  

Dark Energy did not arrive until 1995, when it became increasingly clear that simplest version of inflation + CDM, a flat Universe comprised of 95\% CDM and 5\% baryons didn't work.  There was growing evidence that the  total amount of matter was only about 30\% of critical density \cite{clusterratio}, large-scale structure observations fit, but only if $(\Omega_M\,H_0$/100\,km/sec/Mpc\,$)\simeq 0.2-0.3$, requiring a really small Hubble constant or $\Omega_M \not= 1$, as well as other discrepancies.  The fix which resolved all the problems, as ugly as it seemed, was $\Omega_\Lambda \sim 0.7$ \cite{LambdaCDM}.  Voila!  The first evidence for the acceleration of the Universe was found in 1998 \cite{CA}, and almost overnight,\footnote{My mentor, David Schramm, was supposed to debate Jim Peebles in April 1998 on whether or not $\Omega_0 = 1$.  In December 1997, shortly before he died in a plane crash, he was worried about having to take the $\Omega_0 = 1$ side of the question.  I filled in for David, Peebles was no longer willing to debate $\Omega_0 < 1$ and the title of the debate was changed to, {\it Cosmology Solved?}.  Months after the discovery of cosmic acceleration, $\Lambda$CDM was firmly in place.} $\Lambda$CDM became the new paradigm. Thomas Kuhn would smile.  (Dark energy is the generalization of a smooth, negative pressure component whose equation-of-state $w \equiv p_{DE}/\rho_{DE}$ is close to $-1$ \cite{wCDM}.)


Today, a wealth of cosmological data supports the $\Lambda$CDM model.  It began with the BBN abundance of deuterium that pinned down the baryons density at around 5\% of the critical density, too small to account for the total amount of matter, pointing to the existence of particle dark matter.  In 1992, COBE made the first detection of the anisotropy of the CMB.  Twenty-five years later, the anisotropy has been precisely characterized by the all-sky measurements of the WMAP and Planck satellites and the ground-based measurements down to tiny angular scales of DASI, SPT, ACT and other experiments.  The CMB frontier has moved to polarization and the search for the B-mode signature of inflation produced gravitational waves.

Precision CMB measurements have been and will continue to be crucial to testing the inflationary predictions of almost scale-invariant, nearly-Gaussian curvature perturbations, a spatially flat Universe and an almost scale-invariant spectrum of gravitational waves.  In addition, measurements of the acoustic oscillations imprinted upon the CMB by baryons falling into the dark matter gravitational potential wells has led to unprecedentedly accurate measurements of many important cosmological parameters including the matter and baryon densities, Hubble constant, and the amount of ionized material between us and the surface of last scattering.  

Large redshift-surveys beginning with the SDSS and 2dF surveys have quantitatively characterized the large-scale structure in the Universe.  The large and sophisticated numerical simulations of how structure formed in a $\Lambda$CDM Universe, together with the CMB and LSS results provide strong evidence for both CDM and the gravitational instability theory of structure formation (namely, that structure arose from the gravitational amplification of small density inhomogeneities).

Summing up the second paradigm, $\Lambda$CDM provides a tested account of the Universe from a very early time ($\ll 10^{-6}\,$sec) to the present, some 13.8\,Gyr later:
A very-early period of accelerated expansion driven by the potential energy of a scalar field gave rise to a very-large, smooth, spatially-flat patch that became all that we can see today (and more).  Quantum fluctuations during this accelerated phase grew in size and became the density perturbations that seeded galaxies and other structures.  The conversion of scalar field potential energy into particles produced the quark soup that evolved a baryon asymmetry and long-lived dark-matter particles.  The excess of quarks over antiquarks gave rise to neutrons and protons, and later the nuclei of the lightest chemical elements and finally atoms.  The gravity of the cold (slowly moving) dark matter particles drove the formation of structure from galaxies to superclusters and a mere 5 billion years ago the repulsive gravity of dark energy (at present, indistinguishable from quantum vacuum energy) initiated another period of accelerated expansion.

$\Lambda$CDM has little to say about the beginning or the ending,\footnote{Of course, if the cause of cosmic acceleration is just $\Lambda$, then a gloomy future is easy to predict.} but it does provide sufficient detail about the ``in between" to make the paradigm very testable. Further, $\Lambda$CDM has revealed new physics not contained in the Standard Model: the additional $C$, $CP$ violation needed to produce the baryon asymmetry, dark matter particles, dark energy whose repulsive gravity drives today's accelerated expansion and an early, inflationary phase that lasted for at least 60 or so e-foldings of the scale factor of the Universe.

$\Lambda$CDM has exceeded the expectations of most 1980 cosmologists (certainly this one).  Moreover, its account of the Universe from a very early time when galaxies and all the structures we see today were mere quantum fluctuations to the birth and formation of stars and galaxies is a remarkable achievement.  All of this, with little or no need to discuss or rely upon initial conditions.

However, I suspect that few cosmologists today would simply want to {\it settle} for $\Lambda$CDM  (Success in science is its own worst enemy! -- it breeds ever higher expectations).  Here is a list of some of its shortcomings:
\begin{itemize}
\item{}  $\Lambda$CDM is not a fundamental theory in sense of the Standard Model; it is a highly successful phenomenological model, with some fundamental aspects, e.g., BBN and structure formation
\item{}  What is the dark matter particle?  or is there a dark matter particle?
\item{}  While baryogenesis likely involves neutrino mass and $B+L$ violation arising in the SM, it still remains just an attractive framework with few details.
\item{}  Inflation is still a paradigm, with a Landau-Ginzburg like description of this epoch that so fundamentally shaped the Universe we observe today.
\item{}  Cosmic acceleration and ``the lightness of the quantum vacuum" remain profound mysteries which are likely related.
\item{} What happened before inflation (big bang?) and what is the destiny the Universe?
\item{} And the crazy uncle in the attic of cosmology:  the multiverse.  Born of inflation and the bane of science because of its inability to be tested, it can't stay in the attic forever!

\end{itemize}

\section{Aspirations for the third paradigm}
\noindent {\it What it should address.}
The  third paradigm must aspire to address the big puzzles left unanswered by the second paradigm:  baryogenesis, dark matter, dark energy and inflation.  Or, explain why they are the wrong questions to be asking (not an uncommon occurrence in science).  The issue of the energy of the quantum vacuum and why it is small -- likely related to dark energy and cosmic acceleration is absolutely central as well.  And of course, it would be nice if the third paradigm either illuminated or told us to ignore the multiverse.  Lastly, I am certain that the foundation of the third paradigm will be built upon the deep connections between the very big and very small:  we have yet to plumb the full depths of this profound connection.

\noindent{\it What it need not address.}  Modern cosmology started with stars and galaxies.  $\Lambda$CDM provides the framework necessary to describe the birth, evolution and future of stars, galaxies and large-scale structure.  It does so with such breadth and in such detail that some of the oldest questions of cosmology -- e.g., the origin of the Hubble sequence of galaxies -- can be addressed.  And further, the CDM part of $\Lambda$CDM could be falsified.\footnote{While there continues to be a steady stream of such claims -- dating back to 1983 -- none has yet to convince me that CDM is in trouble. Essentially all such claims involve discrepancies on small scales, where as-of-yet-not-understood hydrodynamics could plausibly resolve the problem.}

I believe cosmology has now bifurcated into two branches: astrophysical cosmology (the story from quarks to us) and fundamental cosmology (the cosmological framework) to the benefit of each. Martin Rees colorfully describes this as the mud wrestlers and the chess players.  The third paradigm should stick with fundamental cosmology and leave astrophysical cosmology, as interesting as it is, to the astrophysicists.

That is not to say that one branch cannot inform the other and probably will.  Right now $\Lambda$CDM is a good enough framework for astrophysical cosmology; when more is learned about dark matter, inflation and dark energy, it may need to be upgraded to get the astrophysical story even more precisely correct.  Conversely, the extraordinary cosmological datasets being produced may provide important observational clues, e.g., about the dark matter particle or evidence that dark energy changes with time, relevant to the cosmological framework.

\noindent {\em How many numbers determine the Universe?}
Sandage had cosmology described by two numbers.  We now know it is more interesting than that.  The Planck fit to the $\Lambda$CDM cosmology has 6 numbers:  the baryon and matter densities; the dimensionless amplitude of the curvature fluctuations; the slight tilt from a scale-invariant spectrum of curvature fluctuations; the sound horizon at last scattering and an astrophysical parameter, the optical depth $\tau$ to the last-scattering surface (determined by the ionization history of the Universe once stars light up).  A fundamental theory of dark matter, dark energy, baryogenesis and inflation would predict all the ``physics" parameters; and a detailed account of the astrophysical evolution of structure formation would predict $\tau$ as well.  Once ``the theory of everything" is known, my aspiration is no additional parameters are needed to explain the Universe.  That is, zero numbers!  

I know of two conjectures that could realize this lofty goal.  In the Hartle-Hawking approach,  the Hamiltonian of the fundamental theory specifies the wavefunction of the Universe.  In the very early days of string theory, Murray Gell-Mann once speculated that the theory of everything is unique.  This would also answer Einstein's famous question:  God didn't have a choice about the laws of physics.\footnote{At the other extreme is, God tried all the possibilities:   The ``marriage" of the multiverse and the infinity of string theory vacua with anthropic reasoning leads to a zoo of universes. We find ourselves in this one because it had laws of physics that permit life to evolve.  Ugh.}

\noindent {\em Initial conditions:  relevant or not?}
For many years, Roger Penrose has taken a nearly opposite view; namely, that initial conditions are everything; in particular, the Weyl curvature.

I am advocating for the opposite, initial conditions are irrelevant.  The Universe is automatic:  the light-element abundances, the baryon asymmetry, the ratio of dark matter-to-baryons, smoothness, flatness, inhomogeneity, ... all arose due to early Universe microphysics.


Sadly, the axion may provide a counterexample to my aspiration.  Here is the dilemma.  The mass of the axion arises
due to the explicit breaking of Peccei-Quinn (PQ) symmetry by small quark masses.  When PQ symmetry spontaneously breaks, at a much larger energy scale, the axion was massless and hence its potential was flat, with no dynamics to determine its value.  In general, the random value of the axion field will be misaligned with the ultimate minimum of its potential.  After chiral symmetry breaking, the axion mass arises and the axion field begins to oscillate with an amplitude set by its misalignment.  These oscillations, which correspond to a condensate of zero-momentum axions, are the dark matter.  The amount of axionic dark matter is determined by the random, initial misalignment, which is only coherent over the size of the horizon at the time of PQ SSB.

 If Peccei-Quinn symmetry breaking occurs after inflation, the axion mass density today is just the statistical average over the random misalignment angles of the many, many horizon-sized regions at the time of PQ SSB that today comprise our Universe.  However, if PQ symmetry breaking occurs first, then our observable Universe resides within one such patch and the amount of axionic dark matter is in essence an initial condition for that patch.
 
That wouldn't be such a big deal except for the fact that its influence could be non-trivial:  if the misalignment is small and the scale of inflation is high, isocurvature fluctuations in the axion would dominate over the usual inflation produced curvature fluctuations, qualitatively changing the nature of structure formation \cite{axionsandinflation}.  The same random misalignment angle determines the ratio of dark matter to baryonic matter, which can have a dramatic impact on the very existence of stars, galaxies and us \cite{axionanthropic}.  This random variable -- an initial condition if you will -- could be important for our Universe if axions are the dark matter.

\noindent {\em How big to think?}
The short answer is, don't think too small!  $\Lambda$CDM provides a strong foundation to speculate from, there are fundamental connections between cosmology and particle physics, and particle physics too has a strong foundation, big aspirations and powerful ideas (David Gross treated us to his overview of them \cite{DavidGross}.)  By the way, theorists have rarely thought too big or taken their ideas too seriously; it is often just the opposite.

In Einstein's big bang theory, the big bang was the singular origin of matter, energy, space and {\it time}, making the question of what happened before the big bang moot.  It seems clear that Einstein did not get the last word on gravity, and if GR's successor (string theory?) cleans up the singularity, the question becomes addressable.  Nonetheless, the {\it emergence} of space, time and the Universe is not only a solution worthy of the question, but also is at the very heart of the connection between particle physics and cosmology.  Where else might one have a better opportunity to address the 400+ year-old question of the fundamental nature of space and time?

On the particle physics side, I see a growing sense that space and time are not fundamental but rather are emergent phenomena, a concept that is difficult to wrap one's head around.   I was struck by David Gross' remark that went something like this, ``space-time is not always the best way to think about things."
As creatures of time, moving in the river of time with an arrow we don't fully understand, I feel better better about not fully understanding his remark while also appreciating the depth of it.  After David's talk, I viewed the ubiquitous WMAP history of the Universe, with its beginning, end and boundaries in a new way.  A cosmic solution can be read in different ways, not just the simple flow in time that a creature of time is most comfortable doing.

\noindent {\em Moving forward.} 
As we have learned from cosmology, the past is hard to predict and the future is even harder. The shift from the mid-second paradigm inflation + CDM to $\Lambda$CDM involved discrepancies.  
Today there is the Hubble tension:  direct measurements of the current expansion rate yield $H_0 = 74\pm 1\,$km/sec/Mpc and ``indirect'' measurements of the expansion rate using CMB anisotropy and the {\it assumption} of $\Lambda$CDM to extrapolate from the early time expansion rate to the present yield $H_0 = 67.5\pm 0.4 \,$km/sec/Mpc \cite{hubbletension}.  The resolution could be a systematic error in one (or both) determinations of $H_0$ or the assumption of $\Lambda$CDM and an indicator of a missing ingredient!

A paradigm begins with its aspirations, is measured by its accomplishments, and the difference between the two defines how revolutionary it was.  How and when we get to the third paradigm remains to be seen.  I am confident that the shift from $\Lambda$CDM to the next paradigm will be at least as revolutionary as that from a cosmology described by two numbers to $\Lambda$CDM and that bold and unsettling new ideas will underpin it. 

\section{The joy of collaborating with Frank}
I end on a personal note, discussing our collaborations over the years.  Frank and I have written 8 papers together; all involved cosmology, and a few were influential.  Some were written when we were together at the ITP in Santa Barbara (before it was the KITP); and two involved work that began at a Nobel Symposium in Graftavellens.  Each was a joyful and insightful experience.  What I remember most vividly was their intensity:  whatever Frank was working on got his full attention, he saw the full possibilities, and that project was -- at least for the moment -- the most important thing he had going.   Here is a quick run down:

\begin{itemize}

\item{}  {\it Reheating an inflationary Universe\,}\cite{RHing}.  In December of 1981, Frank and I received a preprint from Andrei Linde entitled, {\it A new inflationary Universe scenario}, claiming to have cured the problems of Guth's original model of inflation.  The paper looked important, and we worked over the holidays to figure out what was going on.  We wrote down the evolution equation for the scalar field, including the friction effect of the expansion (neglected in Linde's paper), and focussed on how particle production would reheat the Universe.  On a visit to Penn in January, I discovered that Paul Steinhardt was doing similar work as Linde, with his student Andreas Albrecht, and a collaboration was born.  Among other things, the first integration (to my knowledge) of the slow roll equation was done on my HP calculator.

\item{}   {\it Is our vacuum metastable\,}\cite{metastable}? The different symmetry breaking patterns of the $SU(5)$ GUT and how we landed in $SU(3)\times SU(2)\times U(1)$ motivated us, and we explored the possibility that the $SU(3)\times SU(2)\times U(1)$ vacuum was not the ground state, but rather a metastable state with a very-long tunneling time.  The possibility of a transition to a lower energy state has come up almost every time a new accelerator is turned on (Short answer, no.  Cosmic rays have already done the experiment.)

\item{} {\it Formation of structure in an axion-dominated Universe\,}\cite{CDM}. The axion is the quintessential cold dark matter particle, and this is one of the first, if not the first, paper on CDM -- and we have the erratum to prove it.  Starting with the inflationary density-perturbation spectrum and axions, we discussed how structure would form.  It was great fun working with Frank's long-time friend and collaborator, Tony Zee.

\item{} {\it Positron-line radiation as a signature of particle dark matter in the halo\,}\cite{positrons}.  Frank had moved to the IAS by now and I can't remember the origin of this collaboration.  In any case, this paper led to a new signature for dark matter in the halo and became one of the main physics motivations for Sam Ting's AMS experiment.  For a while, AMS had a hint of the positron-line we were talking about.

\item{} {\it Inflationary axion cosmology\,}\cite{axionsandinflation}.  Much of what I said earlier about axions was in this paper that began at the {\it Nobel Symposium on the Birth and early evolution of the Universe} in Graftavallens, Sweden during the summer of 1990.

\item{} {\it Relic gravitational waves and extended inflation\,}\cite{GWs}. This work also began (or was finished) at the Graftavallens symposium.

\item{} {\it Cosmological implications of axinos\,}\cite{axinos}. This was my delightful and productive introduction to Frank's then student and now MIT colleague, Krishna Rajagopal.

\item{}{\it Astrophysics, cosmology and unification of forces.\,}\cite{snowmass}.  This paper, written with two Nobel prizewinners (Frank and Barry Barish), was part of the Snowmass 1994 DPF Planning process. With that author list, it should have been a blockbuster; to date, it has never been cited!
\end{itemize}

Happy birthday Frank and thanks for such a stimulating meeting, one that reflects the breadth and depth of your interests in the big, important ideas in physics.  It has been a joy and privilege to be your friend and colleague for more than 40 years.


\begin{thebibliography}{9}

\bibitem{SWeinberg} S. Weinberg, {\it Gravitation and Cosmology} (J. Wiley \& Sons, 1972).
\bibitem{ASandage} A. Sandage, {\it Physics Today}, {\bf 23}, 34 (1970).
\bibitem{HSTKeyProject} W.L. Freedman et al, {\it Astrophys. J.} {\bf 553}, 47 (2001).
\bibitem{NebenTurner} A.R. Neben and M.S. Turner, {\it Astrophys. J} {\bf 769}, 133 (2013)
\bibitem{WFH} R.V. Wagoner, W.A. Fowler and F. Hoyle, {\it Astrophys. J.}  {\bf 148}, 3 (1967).
\bibitem{Dickeetal} R.H. Dicke, P.J.E. Peebles, P.G. Roll and D.T. Wilkinson, {\it Astrophys. J.} {\bf 142}, 414 (1965).
\bibitem{RubinDM} V. Rubin, {\it Science} {\bf 220}, 1339 (1983).
\bibitem{clusterratio} S.D.M. White et al, {\it Nature} {\bf 366}, 429 (1993).
\bibitem{LambdaCDM} L.M. Krauss and M.S. Turner, {\it Gen. Rel. Grav.} {\bf 27}, 1137 (1995); see also, J.P. Ostriker and P.J. Steinhardt, {\it Nature} {\bf 377}, 600 (1995).
\bibitem{CA} A. Riess et al., {\it Astron. J.} {\bf 116}, 1009 (1998); S. Perlmutter et al., {\it Astrophys. J.} {\bf 517}, 565 (1999).
\bibitem{wCDM} M.S. Turner and M. White, {\it Phys. Rev. D} {\bf 56}, R4439 (1997).
\bibitem{axionsandinflation} M.S. Turner and F. Wilczek, {\it Phys. Rev. Lett.} {\bf 66}, 5 (1991).
\bibitem{axionanthropic} M. Tegmark, A. Aguirre, M.J. Rees, and F. Wilczek, {\it Phys. Rev. D} {\bf 73}, 023505 (2006).
\bibitem{DavidGross} David Gross, in this volume.
\bibitem{hubbletension} E. Di Valentino et al, arXiv:2103.01183v2 (2021)
\bibitem{RHing} A. Albrecht, P.J. Steinhardt, M.S. Turner and F. Wilczek, {\it Phys. Rev. Lett.} {\b 48}, 1437 (1982).
\bibitem{metastable} M.S. Turner and F. Wilczek, {\it Nature} {\bf 298}, 633 (1982).
\bibitem{CDM} M.S. Turner, F. Wilczek and A. Zee, {\it Phys. Lett. B} {\bf 125}, 35 (1983); {\it ibid}, 519 (E).
\bibitem{positrons} M.S. Turner and F. Wilczek, {\it Phys. Rev. D} {\b 42}, 1001 (1990).
\bibitem{GWs} M.S. Turner and F. Wilczek, {\it Phys. Rev. Lett.} {\bf 65}, 3080 (1990).
\bibitem{axinos} K. Rajagopal, M.S. Turner and F. Wilczek, {\it Nucl. Phys. B} {\bf 358}, 447 (1991).
\bibitem{snowmass} B.C. Barish, M.S. Turner and F. Wilczek, {``Astrophysics, cosmology and unification of physics"} in {\it Particle Physics:  perspectives and opportunities (Snowmass 1994)}, edited by R.D. Peccei et al.

\end{thebibliography}
\end{document}